\documentclass{aa}
\usepackage{lineno}

\usepackage{graphicx}
\usepackage[varg]{txfonts}
\usepackage{natbib}
\bibpunct{(}{)}{;}{a}{}{,}
\usepackage{array}
\usepackage{xspace}
\usepackage{lscape}
\usepackage{url}
\usepackage{arydshln}
\usepackage[hyperfootnotes=false, linktocpage=true, breaklinks=true, colorlinks=true, linkcolor=blue, citecolor=blue, urlcolor=blue]{hyperref}
\usepackage[all]{hypcap}
\usepackage{longtable}
\usepackage{afterpage}
\usepackage[dvipsnames]{xcolor}
\usepackage{xfrac}
\usepackage{underscore}
\usepackage{amsmath}
\newcommand{\FeKa}{Fe K\ensuremath{\alpha}\xspace}

\newcommand{\FeKb}{Fe K\ensuremath{\beta}\xspace}

\newcommand{\NH}{\ensuremath{N_{\mathrm{H}}}\xspace}

\newcommand{\vgau}{\xspace{\tt vgau}\xspace}

\newcommand{\xrism}{{XRISM}\xspace}

\newcommand{\spex}{\xspace{\tt SPEX}\xspace}

\newcommand{\ngc}{{NGC~3783}\xspace}

\mathchardef\mhyphen="2D
\begin{document}

\title{Delving into the depths of NGC 3783 with XRISM}
\subtitle{IV. Mapping of the accretion flow with \FeKa emission lines }

\author{
Chen Li \inst{1,2}
\and
Jelle S. Kaastra \inst{2,1}
\and
Liyi Gu \inst{2,1}
\and
Missagh Mehdipour \inst{3}
\and
Megan E. Eckart \inst{4}
\and
Matteo Guainazzi \inst{5}
\and
Erin Kara \inst{6}
\and
Laura~W.~Brenneman \inst{7}
\and
Misaki Mizumoto \inst{8}
\and
Jon Miller \inst{9}
\and
Keigo Fukumura \inst{10}
\and
Ehud Behar \inst{11,6}
\and
Christos Panagiotou \inst{6}
\and
Matilde Signorini \inst{5, 12}
\and
Keqin Zhao \inst{1, 2}
\and
Ralf Ballhausen \inst{13,14,15}
\and
Camille M. Diez \inst{16}
\and
Timothy R. Kallman \inst{14}
\and
Shoji Ogawa \inst{17}
\and
Atsushi Tanimoto \inst{18}
\and
Yoshihiro Ueda \inst{19}
}
\institute{
Leiden Observatory, Leiden University, PO Box 9513, 2300 RA Leiden, the Netherlands  \\ \email{cli@strw.leidenuniv.nl}
\and
SRON Netherlands Institute for Space Research, Niels Bohrweg 4, 2333 CA Leiden, the Netherlands
\and
Space Telescope Science Institute, 3700 San Martin Drive, Baltimore, MD 21218, USA 
\and
Lawrence Livermore National Laboratory, Livermore, CA 94550, USA
\and
ESA European Space Research and Technology Centre (ESTEC), Keplerlaan 1, 2201 AZ, Noordwĳk, the Netherlands
\and
MIT Kavli Institute for Astrophysics and Space Research, Massachusetts Institute of Technology, Cambridge, MA 02139, USA
\and
Center for Astrophysics $|$ Harvard~\&~Smithsonian, 60 Garden St., Cambridge, MA 02138, USA
\and
Science Research Education Unit, University of Teacher Education Fukuoka, Munakata, Fukuoka 811-4192, Japan
\and
Department of Astronomy, University of Michigan, MI 48109, USA
\and
Department of Physics and Astronomy, James Madison University, Harrisonburg, VA 22807, USA
\and
Department of Physics, Technion, Haifa 32000, Israel
\and
INAF - Osservatorio Astrofisico di Arcetri, Largo Enrico Fermi 5, I-50125 Florence, Italy
\and
University of Maryland College Park, Department of Astronomy, College Park, MD 20742, USA
\and
NASA Goddard Space Flight Center (GSFC), Greenbelt, MD 20771, USA
\and
Center for Research and Exploration in Space Science and Technology, NASA GSFC (CRESST II), Greenbelt, MD 20771, USA
\and
ESA European Space Astronomy Centre (ESAC), Camino Bajo del Castillo s/n, 28692 Villanueva de la Cañada, Madrid, Spain
\and
Institute of Space and Astronautical Science (ISAS), Japan Aerospace Exploration Agency (JAXA), Kanagawa 252-5210, Japan
\and
Graduate School of Science and Engineering, Kagoshima University, Kagoshima, 890-8580, Japan
\and
Department of Astronomy, Kyoto University, Kitashirakawa-Oiwake-Cho, Sakyo Kyoto 606-8502, Kinki, Japan
}
\abstract
{
Using XRISM/Resolve $439 \, \rm ks$ time-averaged spectra of the well-known Seyfert-1.5 active galactic nucleus (AGN) in NGC 3783, we investigate the nature of the \FeKa emission line at 6.4 keV, the strongest and most common X-ray line observed in AGN.
Even the narrow component of the line is resolved with evident Fe K$\alpha_{1}$ (6.404 keV) and K$\alpha_{2}$ (6.391 keV) contributions in a 2:1 flux ratio, fully consistent with a neutral gas with negligible bulk velocity. 
The narrow and intermediate-width components have a full-width at half maximum (FWHM) of 350 $\pm$ 50 km/s and $3510 \pm 470 \, \rm km/s$, respectively, suggesting that they arise in the outer disk/torus and/or BLR.
We detect a $10\%$ excess flux around 4 $-$ 7 keV that is not well described by a symmetric Gaussian line, but is consistent with a relativistically broadened emission line. In this paper, we take the simplest approach to model the asymmetric line as a single emission line (assuming either neutral, He-like or H-like iron) convolved with a relativistic disk line model. As expected, the inferred inclination angle is highly sensitive to the assumed ionization state, and ranges between  $i=17-44^{ \circ}$. 
This model also constrains the black hole spin via the extent of the red wing: the required gravitational redshift in the fitted disk-line profile disfavors a non-spinning (Schwarzschild) black hole.
The derived inner radius is close to the radius of the innermost stable circular orbit $r_{\rm ISCO}$ and strongly correlated with the black hole spin.
To better constrain the spin, we fix the inner radius at $r_{\rm ISCO}$ and derive a lower limit on the spin of $a \ge 0.29$ at the 3 $\sigma$ confidence level. 
A Compton shoulder is detected in our data as well as a $2-3 \,  \sigma$ detection of the Cr K$\alpha$ and Ni K$\alpha$ lines.
}
\authorrunning{Chen Li et al.}
\titlerunning{XRISM/Resolve view of the \FeKa emission in NGC 3783}
\maketitle

\nolinenumbers

\section{Introduction}
\label{sect_intro}

The \FeKa line,  at a rest-frame energy of $6.4$ keV, is a prominent emission feature in active galactic nuclei (AGNs) spectra, originating from X-ray fluorescence in dense material surrounding the supermassive black hole (SMBH). Its line width shows significant variation, revealing the diverse physical processes taking place in different regions of the AGN environment (\citealp{Gallo2023}).
Narrow \FeKa lines of a few hundred or thousand km/s likely originate from distant, cold, and dense material such as the torus or outer disk. They provide insights into the structure and geometry of the circumnuclear environment (\citealp{Reynolds_Fabian_1994}; \citealp{Yaqoob2011_Feka}, \citealp{Gandhi2015}). 
In contrast, the broad \FeKa lines, often very extended and skewed with widths of ten to a few hundred thousand km/s, arise from the innermost regions of the accretion disk, where gravitational redshift and relativistic effects significantly broaden and distort the line profile. The presence and characteristics of the broad Fe lines encode valuable information about the disk inner radius, ionization state, the influence of strong gravity, and the spin of the SMBH (\citealp{Fabian1989}; \citealp{Fabian1995}; \citealp{Laor1991}; \citealp{Brenneman2006}).

With the existing X-ray instruments, isolating and characterizing the broad relativistic \FeKa line remains challenging. 
CCDs do not have enough spectral resolution and blend multiple lines, gratings spectra lack the effective area needed for adequate signal-to-noise.
The measurement is critically dependent on several systematic effects, including the shape of the underlying continuum, the presence of "clumpy" absorbers, and potential Fe lines with intermediate line width from the optical/X-ray broad line regions. To overcome these challenges, together with high signal-to-noise data, high-resolution spectroscopy in the \FeKa band coupled with simultaneous modelling of all the relevant spectral components are essential especially in time-averaged spectra \citep{Reynolds2012}.
Energy-dependent lag spectra can further test whether the broadband variability around the \FeKa band originates from clumpy absorption or is intrinsic to the central source (\citealp{Zoghbi2019}; \citealp{Reis2012}).

The recent launch of the X-Ray Imaging and Spectroscopy Mission (XRISM) marks a significant advancement in our ability to study these Fe lines \citep{Tashiro2025}. 
With its high spectral resolution, the Resolve instrument onboard XRISM can distinctly separate emission and absorption components with varying widths (\citealp{Ishisaki_2025}; \citealp{Kelley2025}). Early results demonstrate that what was previously thought to be a single narrow line is actually a complex blend of several components, each originating from different regions such as the torus, the broad line region, and the inner accretion disk \citep{XRISM_collaboration2024}.

NGC 3783 (z=0.00973; \citealp{Theureau1998}) is a well-studied Seyfert 1 galaxy, extensively examined for its X-ray wind components (\citealp{Kaspi2002}; \citealp{Netzer2003}; \citealp{Mao2019}; \citealp{Gu2023}). Due to its brightness in the Fe K band, it was selected as a XRISM performance verification (PV) target to probe the highly ionized gas resolved by velocity \citep{Mehdipour2025}.
Further the full absorption measure distribution of the X-ray outflow will be determined by combining with XMM/Newton reflection grating spectrometer (RGS) observations (Zhao et al. 2025 in prep, paper V).
Simultaneously, it offers an unprecedented opportunity to investigate the nature of the \FeKa emission line with the help of the resolved Fe K band absorption lines with Resolve, which is the main scientific goal of our present paper.

Using the combined data taken during $2000 - 2001$ by the Chandra High Energy Transmission Grating Spectrometer (HETGS), the full width at half maximum (FWHM) of the narrow \FeKa line core in NGC 3783 is approximately 1,800 km/s, located between the BLR and NLR (\citealp{Kaspi2002}; \citealp{Yaqoob2005}).
It was also found that the excess flux around the base of the narrow \FeKa line core can be modeled with either a Compton-scattering ‘‘shoulder’’ or a relativistic line from the inner disk with low inclination angle of $11^\circ$ or less. 
With the XMM-Newton EPIC data taken in 2001, \cite{Reeves2004} have reported not only a strong broader \FeKa line at $6.4$ keV, but also another peak feature at $7.05$ keV probably originating from a blend of the neutral \FeKb line and the hydrogen-like line of Fe at $6.97$ keV. 
They also reported a weak, broad red wing, attributed either to an ionized warm absorber or to relativistic effects (\citealp{Reeves2004}).

Using a deep (210 ks) Suzaku observation in 2009, \cite{Brenneman2011} and \cite{Reis2012} reported a strongly broadened and skewed iron line and concluded that the black hole is rapidly spinning in the prograde sense.
On the other hand, \cite{Patrick2011} obtained different results using the same data set, preferring a slowly or
retrograde spinning black hole $a \le - 0.04 $.
Motivated by the discrepancy above, \cite{Reynolds2012} performed a Monte Carlo Markov Chain based investigation of black hole spin in NGC 3783, and identified a (partial) modelling degeneracy between the iron abundance of the disk and the black hole spin parameter.
Moreover, two different methods used by \cite{Capellupo2017}, namely relativistic reflection and continuum fitting, yield a wide range for the spin parameter, albeit with a high probability for a near-maximal spin if a disk wind is included in the reflection model.

Using all the archival data collected by the Chandra/HETGS, \cite{Danehkar2025} pointed out the relativistically broadened iron emission is consistently associated with a near-maximal black hole spin across all analysed datasets. This high spin is a constant finding even when the system transitions to different spectral states, such as the hard state due to some transient obscuration events caused by eclipsing outflow material near the X-ray source \citep{Mehdipour2017}, which was present in data from 2013 to 2016.
The analysis of HETGS data by \cite{Danehkar2025} also showed that the narrow \FeKa line from distant regions has a few hundred km/s excess redshift velocity relative to the rest frame, suggesting a warped disk scenario causing the far side of the disk was more illuminated.

With its unprecedented energy resolution and enhanced effective area in the Fe K band, the XRISM Resolve instrument is poised to advance our understanding of the \FeKa complex in NGC 3783.
This paper is organized as follows. 
Section \ref{sect_model} presents the spectral modelling method and analysis. 
Section \ref{result} describes the narrow \FeKa observational characters and the broad iron emission spectral modelling results. 
We summarize and discuss all of results in Section \ref{sect_discuss}.

\section{Spectral modelling and analysis}
\label{sect_model}

NGC 3783 was observed by XRISM/Resolve during $10$ days (OBSID 300050010) from July $18-27$ in  2024. 
We use exactly the same spectral data as paper I and \cite{XRISM_Collaoration_2025_Kaastra}, we refer readers to these works for comprehensive details on data reduction and analysis.
For clarity, we highlight the related information in our current paper.
Of the 36 pixels of Resolve instrument, pixel 12 (calibration pixel) and pixel 27 (which shows unpredictable gain jumps) were excluded from the analysis. Only high-resolution primary (Hp) events (\citealp{Ishisaki2018}) were selected for our analysis. The count rate was approximately 0.1 s$^{-1}$ pix$^{-1}$ in the four central pixels and ranged from 0.001 to 0.1 s$^{-1}$ pix$^{-1}$ in the outer pixels. 
To mitigate contamination, all low-resolution secondary events were excluded before creating the extra-large response matrix file (RMF), which includes all known instrument effects.

The gain uncertainty is 0.16 eV at 5.9 keV and the systematic energy scale uncertainty is 0.3 eV in the band of 5.4 $-$ 8.0 keV, which are uncorrelated and can be root-sum-squared.
The total systematic uncertainty corresponds to 15 km s$^{-1}$ at 7 keV.
The FWHM of 5 eV at 6.4 keV is equivalent to 234 km /s at 6.4 keV. 
The non-X-ray background (NXB) spectrum was modeled from the Resolve night-Earth data database\footnote[1]{https://heasarc.gsfc.nasa.gov/docs/xrism/analysis/nxb/index.html}, while the sky background  was ignored because its contribution above 2 keV is negligible.

We modelled the time-averaged $439\, \rm ks$ \xrism/Resolve spectrum (optimally binned to $\approx 2 \, \rm eV$ with {\tt obin}, see \citealp{Kaastra_Bleeker2016}) using \spex\ {\tt v3.08.01} \citep{Kaastra1996, Kaastra2024} along with an atomic database SPEXACT version 2.07.00. 
Spectral fitting was performed with C-statistics \citep{Kaastra2017}. 
The cosmological redshift was fixed at 0.009730 \citep{Theureau1998}. Galactic X-ray absorption was calculated using the {\tt hot} model in \spex, with a column density of $\NH = 9.59 \times 10^{20}$~cm$^{-2}$ \citep{Murphy1996}. The abundances of all components were set to the proto-solar values of \citet{Lodders2009}.

\subsection{Modeling of the photoabsorption continuum}
\begin{figure}[!tbp]
\centering
\hspace*{-0.1cm}\resizebox{\hsize}{!}{\includegraphics[angle=0]{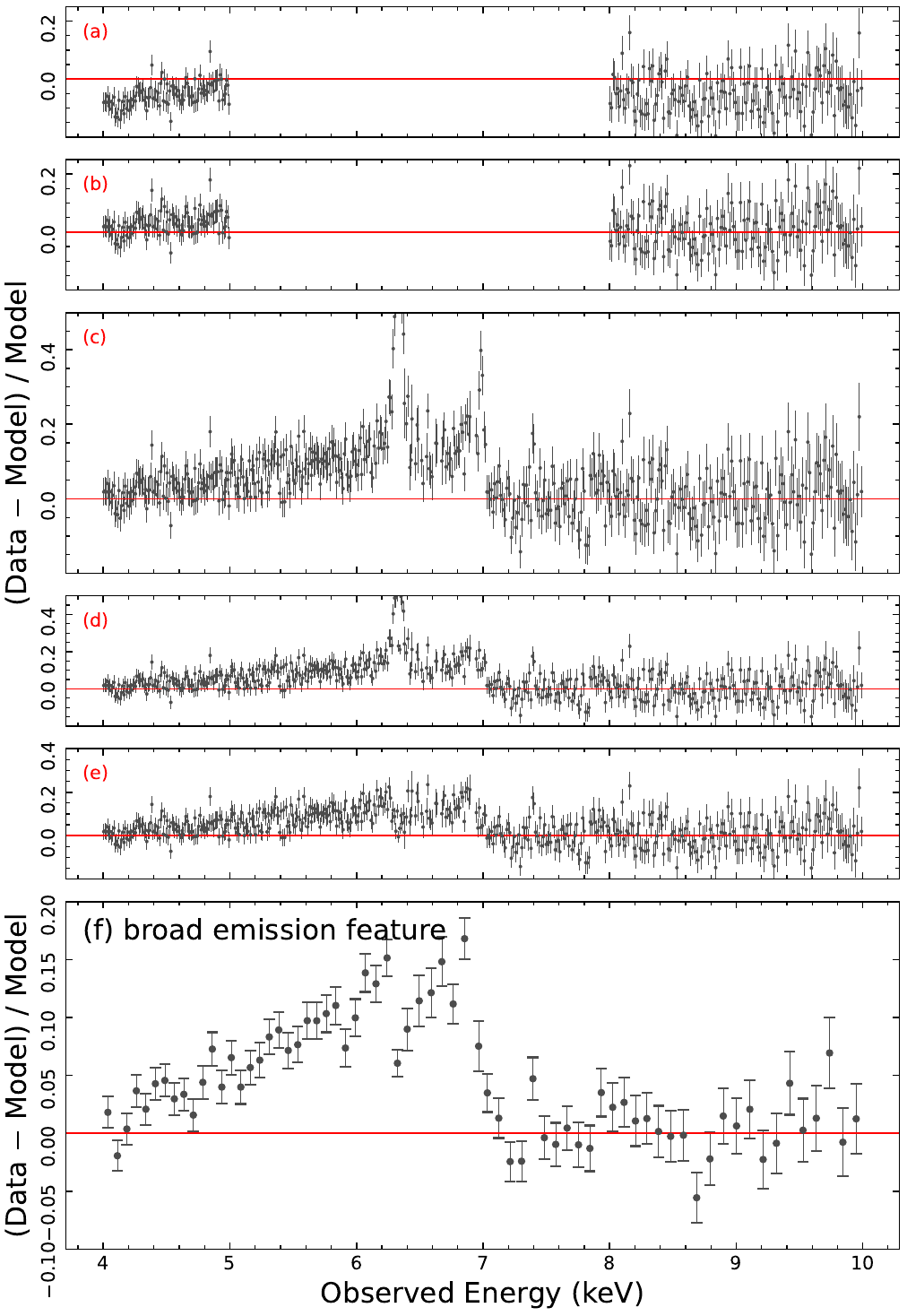}}
\caption{XRISM/Resolve residual spectra for each added model component one by one: (a) power-law; (b) photoabsorbed power-law; (c) all data, with narrow absorption lines excluded as described in the text; (d) narrow Fe core added; (e) intermediate neutral Fe component added; (f) residual broad emission feature, re-binned to 60 eV to better reveal the broad emission structure.
}
\label{fig_chi}
\end{figure}

The time-averaged Resolve spectrum was modeled by a power-law component in the 4$-$10 keV band, excluding the 2$-$4 keV as well as the 5$-$8 keV ranges where multiple absorption line features are present (\citealp{Kaspi2001}). As shown in Figure \ref{fig_chi} panel (a), the model falls below the data, in particular towards 4 keV,  suggesting the presence of an additional underlying component.

Building on the models reported by paper I and paper V, our second step is to incorporate warm absorber components to model in particular the continuum absorption.
We apply 5 {\tt slab} components to incorporate the impact of photoabsorption which adopt the column density of ions which are taken from paper V. 
These 5 components in order correspond to components C1, A1, A3+B+C2, and A2, of paper V as described from a joint fit using {\tt pion} components of the combined Resolve and RGS spectra.
The {\tt pion} model is a fully self-consistent model that computes the absorption and emission for a photoionized plasma (\citealp{Mehdipour2016}; \citealp{Miller2015}). The {\tt slab} model is a pure empirical model for the transmission of a layer of material with arbitrary ionic column densities (\citealp{Kaastra2002}).
To maintain physical consistency, we adopt the ionic column densities obtained from their {\tt pion} fits as inputs to {\tt slab} in our current work, ensuring that the absorption continuum is treated in a physically consistent manner and reduces computational cost.

\begin{figure}[!tbp]
\centering
\hspace*{-0.05cm}\resizebox{\hsize}{!}{\includegraphics[angle=0]{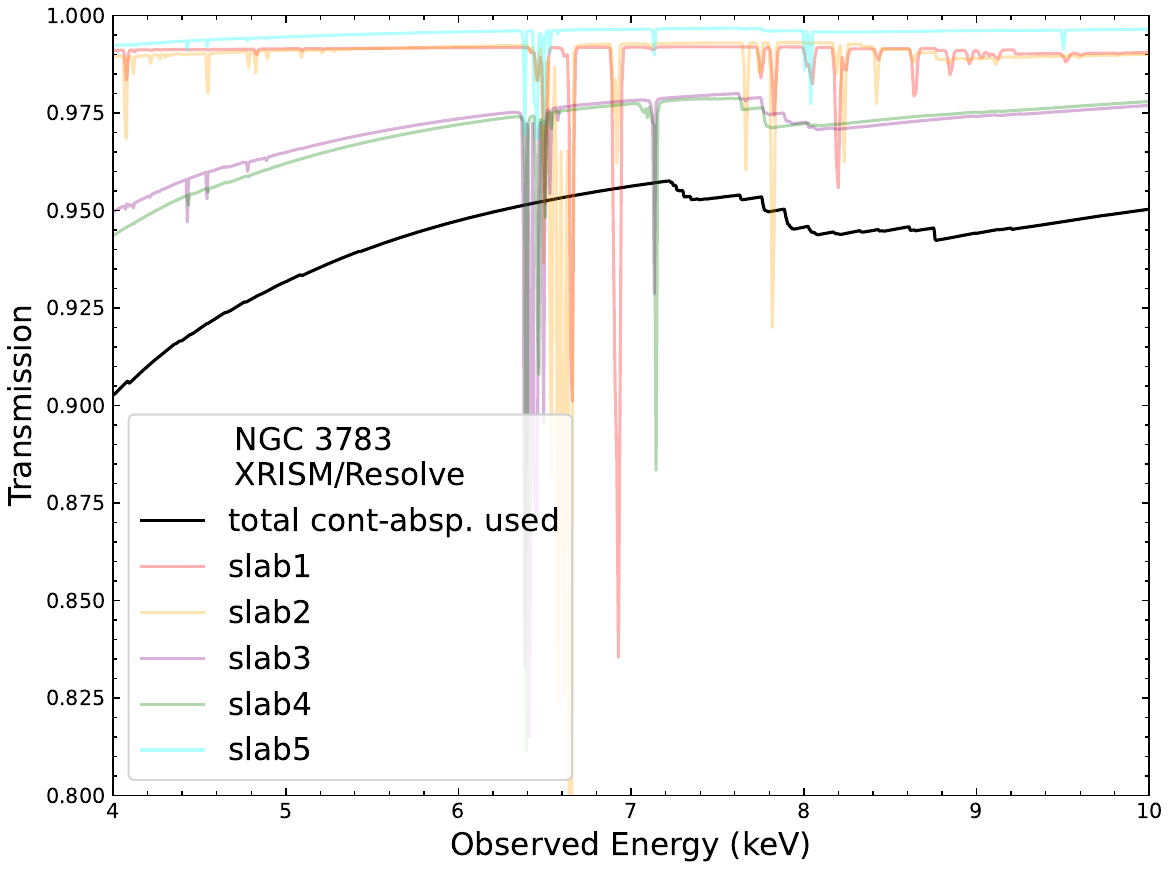}}\vspace{-0.1cm}
\caption{The transmission of five warm absorber components using {\tt slab} model with Resolve data.
In our current emission-line and continuum modelling, we include only the total continuum absorption from the warm absorbers (shown in black).
}
\label{fig_slab_transm}
\end{figure}

The transmission of these components is plotted in Figure \ref{fig_slab_transm} for reference.
We only retain the total continuum absorption (black curve of Fig. \ref{fig_slab_transm}) caused by these warm absorbers when we model the continuum using the broad band data of $4-5$ keV and $8 -10 $ keV.
Including the weak lines in these bands has a negligible effect on the derived power law parameters.
Panel (b) of Figure \ref{fig_chi} illustrates the improvement in the model by incorporating this continuum absorption. 
We did not incorporate the component X, the fast ($v$ $\sim$ 14, 300 km/s) outflowing wind component in paper I, as this component only has 1$\%$ $-$ 2$\%$ effect on the continuum seen from the transmission.
We have not included the Compton hump in our continuum model. Based on the work of \cite{Mao2019}, its contribution is negligible at 10 keV (typically less than $1\%$) and is even smaller at lower energies.

To minimize systematic uncertainties in modeling Fe-K emission components, we excluded all spectral bins from the spectral analysis, which contained significant absorption lines in the 6.38 – 7.1 keV range (see Fig. 1 of Paper I).
Panel (c) of Figure \ref{fig_chi} shows the residuals of the full spectrum (excluding the absorption line data mentioned above) to the continuum model described in this section.
And the full spectrum was applied to the following data analysis.

\subsection{Narrow- and intermediate-width Fe K emission modelling}\label{narrow} 

The neutral Fe K$\alpha$ emission is not a single emission line, but a line complex comprised of emission from numerous atomic transitions. An accurate model of the natural line shape of the \FeKa complex must be incorporated when the velocity broadening is less than $\sim 2000$ km/s (\citealp{Yaqoob2024}).
Here, we adopt the line profile of the  seven-Lorentzian decomposition derived from high-precision laboratory measurements reported by \cite{Holzer1997_Kab}. For \FeKa, the profile provided by \cite{Holzer1997_Kab} is dominated by two strong complex, Fe K$\alpha_1$ ($6.404 \, \rm keV$) and Fe K$\alpha_2$ ($6.391 \, \rm keV$), separated by about 13 eV. 
This split of Fe K$\alpha_1$ and Fe K$\alpha_2$ peaks is seen in the time-averaged Resolve spectrum of NGC 3783.  
We adopt the \FeKb line profile from the same laboratory measurement, scaling the flux ratio between \FeKa and \FeKb to 8:1, which is the characteristic ratio for neutral Fe fluorescence (\citealp{Yamaguchi2014}).
Finally, we combine them into a {\tt file} model in SPEX, which includes Fe K$\alpha_1$,  Fe K$\alpha_2$, and \FeKb lines convolved with the same velocity broadening and redshift for representing the cold neutral line.

Similarly to the case of NGC 4151, the apparently narrow peak of the Fe K$\alpha$ line complex 
may contain different velocity broadening
components (\citealp{XRISM_collaboration2024}). 
To investigate this, we employ a multiple component model. Each component consists of a fixed intrinsic line profile as described above, which is convolved with a Gaussian broadening and shifted by a free velocity.
Two components are found to be sufficient to reproduce the observed narrow Fe emission feature. 
Panel (d) of Figure \ref{fig_chi} shows the remaining residuals after inclusion of a narrow emission core with a few hundred km/s.
Panel (e) shows the residuals after including the second intermediate width component of a few thousand km/s.

\subsection{Broad emission feature between $4 - 7$ keV}\label{model_bFe}
After modelling the continuum and the narrow emission lines, a broad, skewed residual remains, reaching $\sim 10\%$ above the model between 4 – 7 keV (Fig.~\ref{fig_chi}, panel e). In panel (f), we rebinned this asymmetric structure to 60 eV bins to better highlight the broad emission.
Such a broad emission feature can potentially arise from a relativistically broadened iron line of the inner accretion disk.
Under this assumption, we fit the structure using the unbinned high resolution spectrum ($\sim$ 2 eV) assuming an intrinsically narrow iron line convolved with a relativistic line broadening model, {\tt spei} in {\tt SPEX} (\citealp{Speith1995}). 
This kernel is designed to handle a variety of black hole spins, inclination angles of disk, and complicated emissivity profiles, as well as different regions of the disk.
We used a {\tt delt} line model convolved with {\tt spei} to model the broad emission feature. 
For the {\tt delt} component, the normalization parameter is free and the energy of the line is kept frozen at a specified value.
For the {\tt spei} component, we allow the inner radius of the disk ($\rm r_{in}$), black hole spin ($a$), inclination angle ($i$), and emissivity slope ($q$) to vary freely during the fit. 
We fix the emissivity scale height parameter 
$h$ in the {\tt spei} model to zero, corresponding to a pure power-law emissivity profile \footnote[2]{https://spex-xray.github.io/spex-help/models/speith.html}.  
Although a common empirical approach is to fix the inner radius $\rm r_{in}$ to $\rm r_{isco}$, the theoretical innermost stable circular orbit (ISCO) which depends on the spin, here we permit the inner radius to extend beyond the ISCO to investigate the possibility of disk truncation outside the ISCO.

The broad iron line originates from reflection off the inner regions of the accretion disk. The dominant iron ion species contributing to this feature depend on the ionization state of the disk surface.
To explore different ionization states of the inner disk, we investigate limiting cases where the intrinsic lines are either neutral or highly ionized. For the neutral line scenario Model A, we include two delta line components with fixed energies, one for \FeKa at $6.4 \, \rm keV$ and another one for \FeKb at $7.05 \, \rm keV$, coupling the normalization of \FeKb to \FeKa and keeping the flux ratio between \FeKa and \FeKb 8:1, 
both are convolved by the same {\tt spei} model. 
For the highly ionized disk cases, we explore the He-like Model B and H-like Model C ionization conditions. In these scenarios, we used the {\tt delt} model and fixed the line centres at $6.7 \, \rm keV$ and $6.97 \, \rm keV$ respectively, convolved by the corresponding {\tt spei} model. 
For all models the line flux is a free parameter.
Adding the broad component to the fit model A, B and C, it improves the fit C-statistic by 1044, 1038 and 1033, respectively.

\subsection{Compton shoulder modelling in $6.15-6.30$ keV}
\begin{figure}[!tbp]
\centering
\hspace*{-0.05cm}\resizebox{1.0\hsize}{!}{\includegraphics[angle=0]{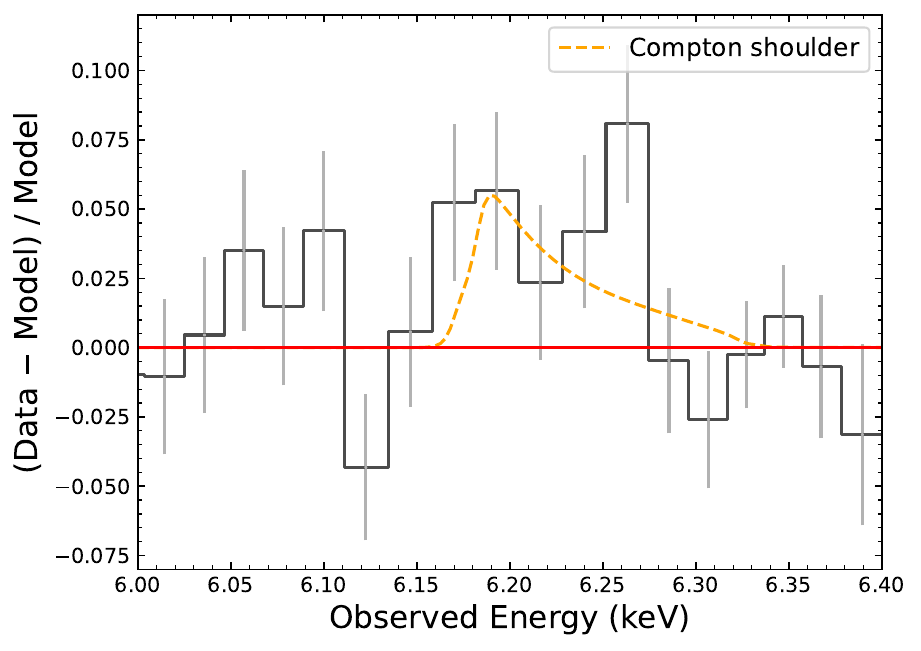}}\vspace{-0.01cm}
\caption{XRISM/Resolve residual plot of NGC 3783 after modelling the \FeKa core with three emission components. 
The residuals are modelled with the yellow line in the Figure \ref{fig_spec}.
}
\label{fig_cs_chi}
\end{figure}

The above components adequately describe the observed Resolve spectrum, with only a potential excess remaining between 6.0 and 6.4 keV (Figure \ref{fig_cs_chi}). Although the excess has a significance of only 2.1$-$2.4 $\sigma$, the tail-like spectral feature could be generated via Compton scattering of neutral matter, known as a Compton shoulder (CS) \citep{Odaka2016}.
We modeled this feature with the {\tt file} model for neutral intrinsic \FeKa and \FeKb profiles of the emission line (see Sect. \ref{sect_model}) convolved with the {\tt vcom} model.
The {\tt vcom} model is a phenomenological model for single Compton scattering, with an option for an empirical skewness correction $a$ such that the profile is proportional to (1+$y^2$)(1+$ay$) with $y$ the scaled energy between -1 and 1.
In the {\tt vcom} model, we keep the skewness parameter ($a$) thawed to fit the data.
Including the CS model improves the C-stat value by $12$, $9$ and $9$ respectively, for model A, B and C.

\subsection{Neutral Cr K$\alpha$, Ni K$\alpha$} 
There are still a few line-like excesses at the Cr K$\alpha$ (5.36 keV) and Ni~K$\alpha$ (7.40 keV) energies. Similarly to Fe, the Cr K$\alpha$ and Ni K$\alpha$ lines are modeled using natural line shapes provided by \cite{Holzer1997_Kab}, and convolved with a Gaussian broadening profile (\vgau). We force the broadening of these two lines to be equal to that of the narrow \FeKa component.
The Cr K$\alpha$ component improves the C-stat by 8, 7 and 12 for Model A, B and C, respectively.
The Ni~K$\alpha$ component improves the C-stat by 11, 15 and 15 for Model A, B and C, respectively.

\subsection{Other ionized emission lines}
The $6.8 - 7.1$ keV band of Resolve is complex with possible contributions of neutral \FeKb, \ion{Fe}{XXVI} Ly$\alpha$ absorption and emission, the \ion{Fe}{I} K absorption edge, and the relativistic iron line blue edge.
The weak \ion{Fe}{XXVI} emission line has only less than 2$\sigma$ significance after modeling.
Due to its weakness and the complexity of this narrow energy band, we refrain from a more detailed analysis.
In addition to weak \ion{Fe}{XXVI} emission lines, other weak relative low ionized iron emission lines in the 6.4 $-$ 6.6 keV band could possibly be present.
We did not investigate them further because they are not significant in our current data.

\begin{figure*}
\centering
\includegraphics[width=17cm]{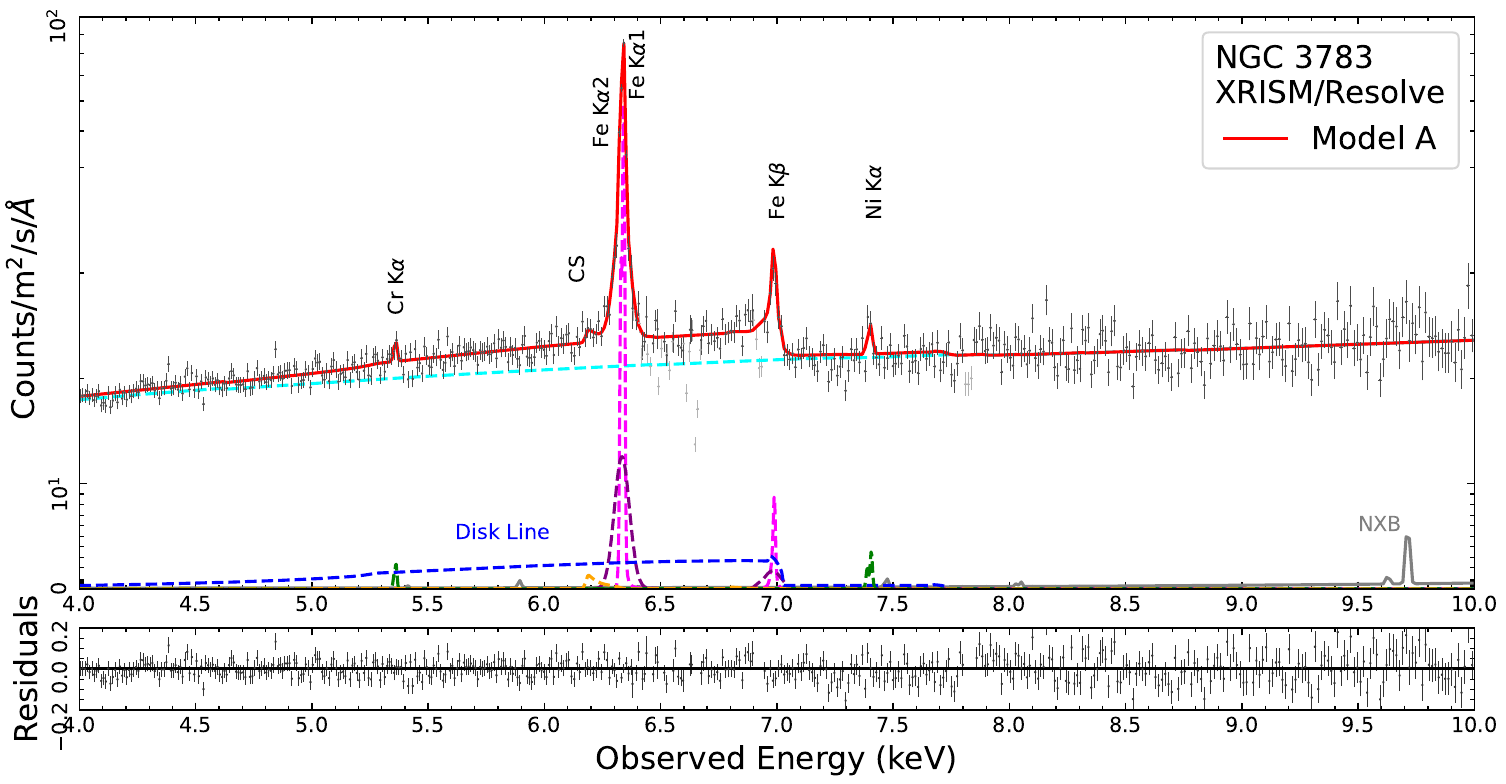}
\includegraphics[width=17cm]{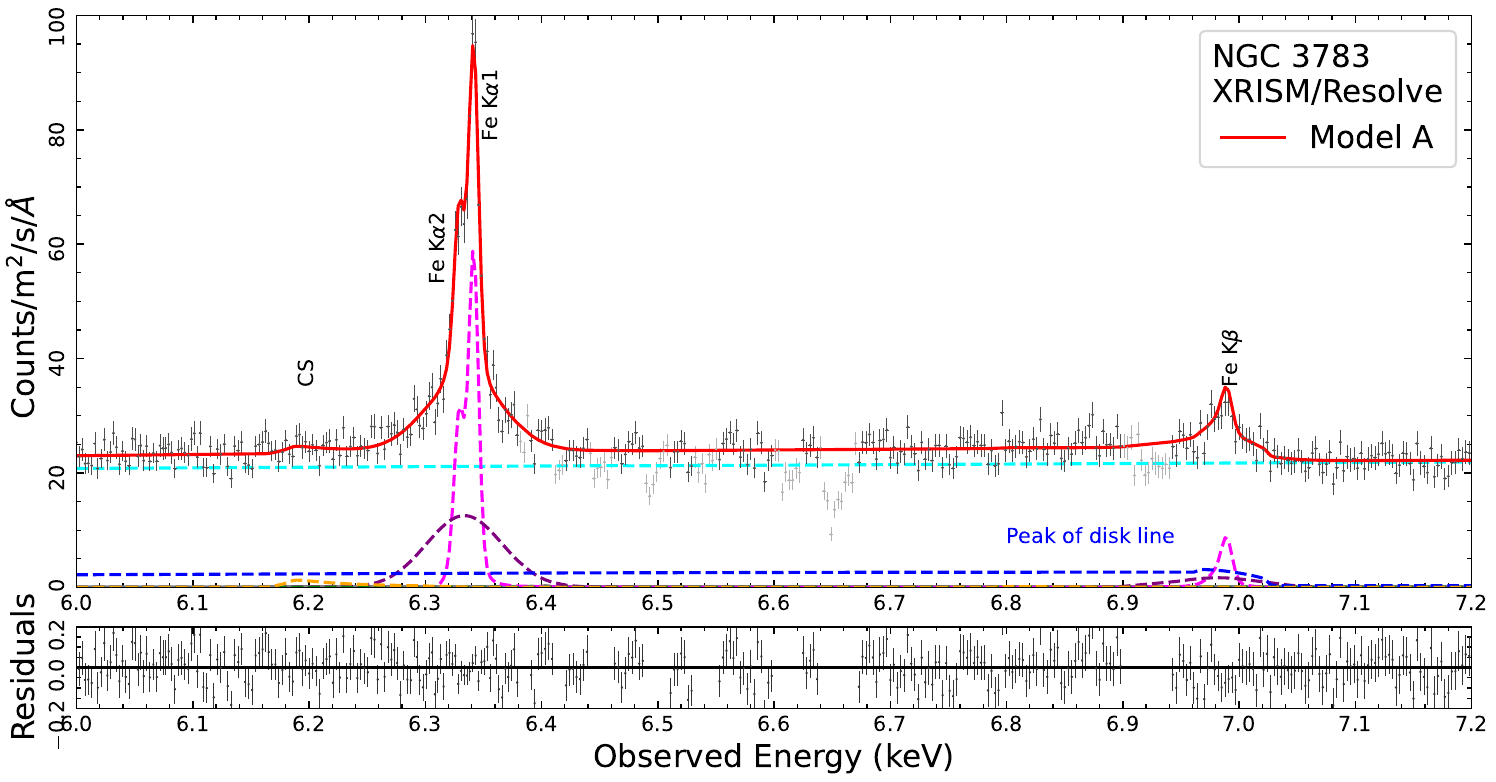}
\caption{XRISM/Resolve time-averaged spectrum of \ngc with the best-fitting model A shown as red solid line.  
The top two panels show the Resolve background-subtracted spectrum with individual model components and the corresponding fit residuals. 
The fit residuals are defined as (data $-$ model) / model.
In our fitting, we used all data points in black excluding the strong absorption line data points (in gray) for modelling continuum and emission lines independently.
For a clarity of display the spectrum in the top panel, it is additionally binned up to 8 eV. 
The bottom two panels provide a close-up view of the Fe K band and its fit residuals at the model optimal bin size of 2 eV. 
In both panels, the strongest emission features are labelled. The individual components are plotted by dashed lines representing the power law (cyan), iron narrowest (magenta)  and intermediate (purple) component, relativistic iron line (blue), Compton shoulder (orange), Cr K$\alpha$ and Ni K$\alpha$ (green). The background model is depicted by the gray curve.}
\label{fig_spec}
\end{figure*}

\begin{figure}[!tbp]
\centering
\hspace*{-0.1cm}\resizebox{\hsize}{!}{\includegraphics[angle=0]{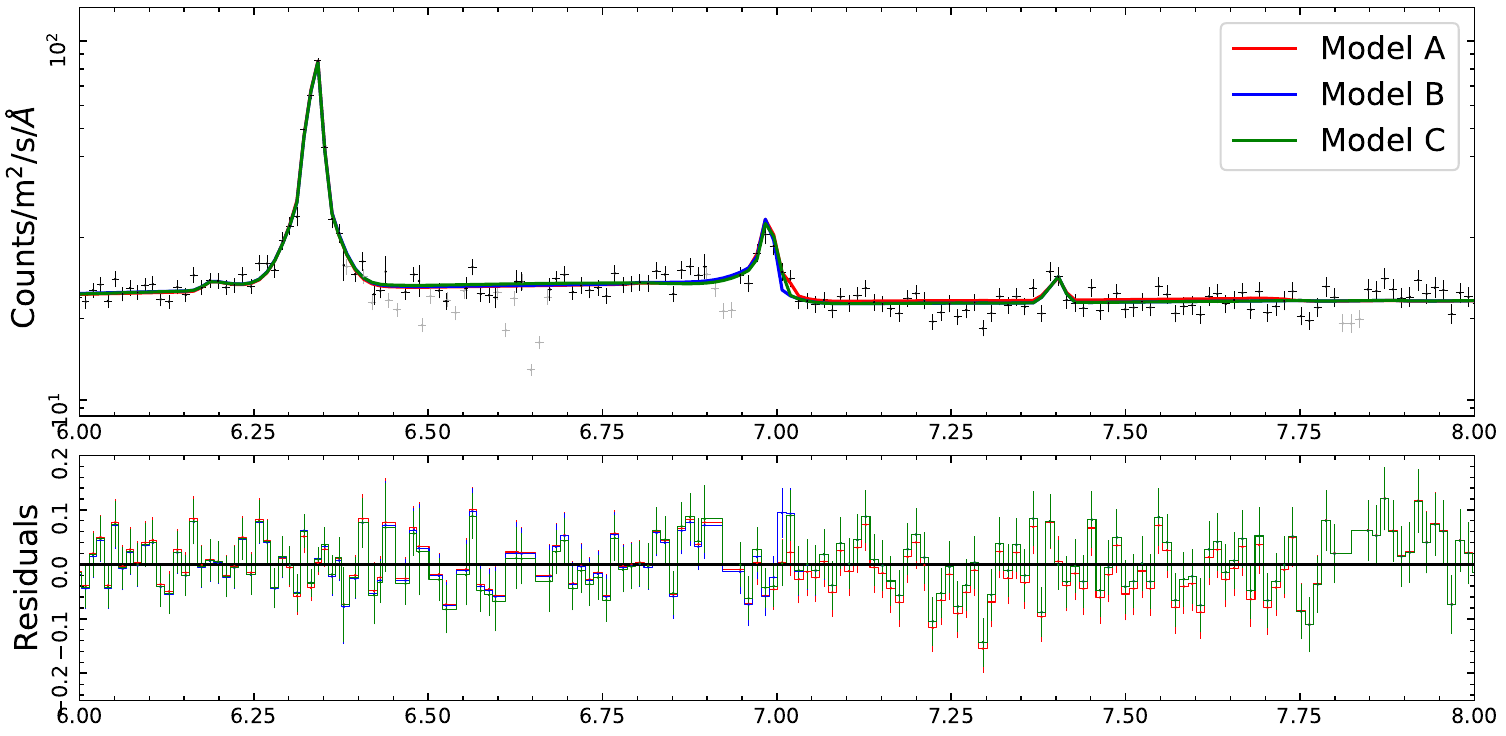}}
\hspace*{-0.1cm}\resizebox{\hsize}{!}{\includegraphics[angle=0]{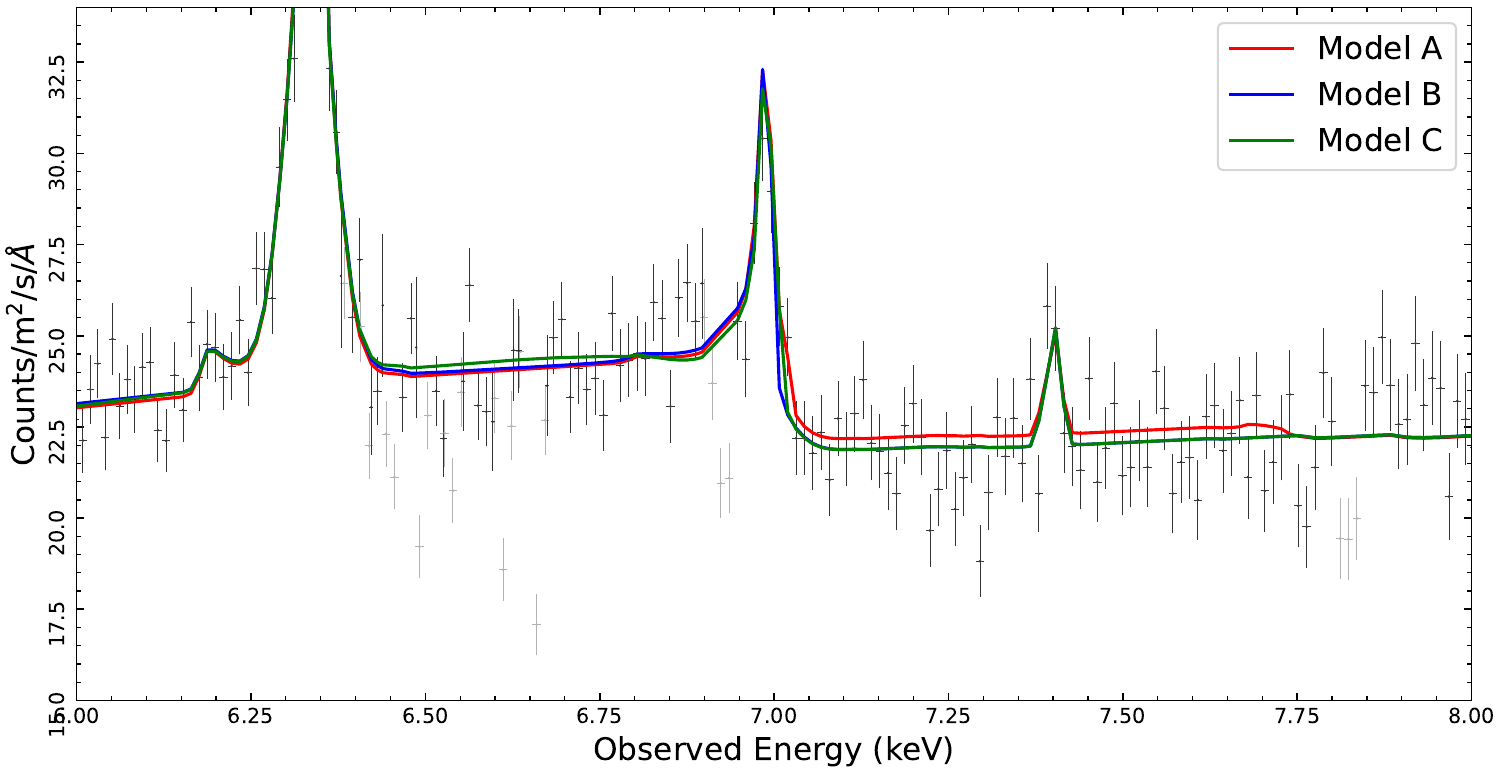}}
\caption{XRISM/Resolve spectra of NGC 3783 fitted with three different models: Model A (red), Model B (blue), and Model C (green). The top two panels illustrate the subtle differences among these models. Residuals are calculated as (data $-$ model) / model, similar to Fig. \ref{fig_spec}. The bottom panel shows a close-up view of the complex spectral structure in the 6.75 $-$ 7.1 keV energy range.
Spectral bins containing narrow absorption features are included in the spectral plot but excluded from the residual plot.
}
\label{fig_chi_comparison}
\end{figure}

%
\renewcommand{\arraystretch}{1.5}  
\begin{table*}\label{tab_best-fit-par}
\begin{center}
\caption[]{Best-fit parameters of the \xrism Resolve spectrum of \ngc fitted between $4-10$ keV for the three models A, B and C as described in the text.
}
\footnotesize
\begin{tabular}{llllll}
\hline\hline\noalign{\smallskip}
  Component  & {\tt SPEX} acronym & Parameter (units)	   & Model A  & Model B	   & Model C         \\
\noalign{\smallskip}
\hline
Galactic column  & {\tt hot}  & $N_{\rm H}$ ($10^{24} \, \rm m^{-2}$)  & $9.59\tablefootmark{a}$   &$9.59\tablefootmark{a}$   &$9.59\tablefootmark{a}$  \\
\hline
Power law        & {\tt pow}  & $\Gamma$           &	$	1.76	\pm	0.01	$	&	$	1.76	\pm	0.01	$	&	$	1.76	\pm	0.01	$ \\
                 &            & Norm ($10^{49} \, \rm ph \, \mathrm{s}^{-1} \, \mathrm{keV}^{-1} $)  &	$	369	\pm	6	$	&	$	369	\pm	6	$	&	$	369	\pm	6$          \\
                 &            & Luminosity ($10^{36}$ \text{W})           &	$	1.38	\pm	0.02	$	&	$	1.38	\pm	0.02	$	&	$	1.38	\pm	0.02	$  \\
\hline 
Warm absorbers    & {5 $\times$ \tt slab}(s)    &   &see Figure \ref{fig_slab_transm}  &   & \\
\hline 
$\rm Fe \, K\alpha \beta_{nar}$  & {\tt file\tablefootmark{b}{}}  & nr of Ph ($10^{48} \, \rm ph s^{-1}$)  & $7.7\pm	0.5$  & $7.8\pm	0.8$  &$7.8\pm	0.7$  \\
                            & {\vgau} & $\sigma$ ($\rm km/s$)   &	$	150	\pm	20	$	&	$	140	\pm	20	$	&	$	140	\pm	30	$     \\
                            & {\tt reds}     &$\mathrm{c}*z$ \, ($\mathrm{km/s}$)      &	$	6	\pm	10	$	&	$	5	\pm	10	$	&	$	5	\pm	10	$   \\
                            &           & Luminosity ($10^{33}$ \text{W})  &	$	7.8	\pm	0.5	$	&	$	7.7	\pm	1.1	$	&	$	7.7	\pm	1.5	$   \\
\hline
$\rm Fe \, K\alpha \beta_{intm}$  & {\tt file\tablefootmark{b}}  & nr of Ph ($10^{48} \, \rm ph s^{-1}$)  &	$	7.5	\pm	0.6	$	&	$	7.4	\pm	0.7	$	&	$	7.4	\pm	0.9	$   \\

                     & {\vgau}  & $\sigma$ ($\rm km/s$)  &	$	1490	\pm	200	$	&	$	1510\pm	210	$	&	$	1490\pm	290	$     \\
                     & {\tt reds}   &$\mathrm{c}*z \, (\mathrm{km/s})$  &	$210\pm90$	&
                     $	190	\pm	90	$	&	$	210	\pm	90	$    \\
                     &     & Luminosity ($10^{33}$ \text{W})    &	$	7.8	\pm	0.7	$	&	$	7.6	\pm	1.0	$	&	$	7.7	\pm	1.5	$ \\
\hline
%
& & ionization state
  & $\ion{Fe}{II}\ \mathrm{K}\alpha,\,\mathrm{K}\beta$
  & $\ion{Fe}{XXV}\ \mathrm{He}\,\alpha$
  & $\ion{Fe}{XXVI}\ \mathrm{Ly}\,\alpha$ \\
Fe K$\alpha\ \beta_{\mathrm{brd}}$  & {$\tt delt$}  & Norm ($10^{49} \, \rm ph \, \mathrm{s}^{-1} $)  &	$	3.8	\pm	0.2	$	&	$	4.0	\pm	0.2	$	&	$	3.8	\pm	0.2	$  \\
                   &           & $E$ (keV)   & $ 6.4\tablefootmark{a}, 7.05\tablefootmark{a} $     & $ 6.7\tablefootmark{a} $    & $ 6.97\tablefootmark{a} $  \\
                  & {\tt spei}  & $\rm r_{in}$ ($\mathrm{GM/c^{2}}$)   & $2.69^{+0.27}_{-0.00}\tablefootmark{c}$    & $3.47^{+0.20}_{-0.30}\tablefootmark{c}$     & $ 3.69^{+0.22}_{-0.33}\tablefootmark{c}$   \\
                  &             & $\rm r_{out}$ ($\mathrm{GM/c^{2}}$)  & $400\tablefootmark{a}$     & $400\tablefootmark{a}$    & $400\tablefootmark{a}$   \\
                  &             & $a$   & $0.84^{+0.06}_{-0.01}\tablefootmark{c}$    & $ 0.75^{+0.04}_{-0.10}\tablefootmark{c}$     & $ 0.80^{+0.05}_{-0.21}\tablefootmark{c}$  \\
                  &             & $\rm r_{isco}$ ($\mathrm{GM/c^{2}}$) & $2.69$     & $3.15$    & $2.93$   \\
                  &            & $ i $ (deg)  &	$	44.3	\pm	0.2	$	&	$	32.3	\pm	1.5	$	&	$	17.6	\pm	1.4	$ \\
                  &            & $q$ (emissivity slope)  &	$	4.0	\pm	0.2	$	&	$	3.6	\pm	0.3	$	&	$	3.3	\pm	0.1	$  \\
                  &            & $h$ (emissivity scale height)  & $0\tablefootmark{a}$    & $0\tablefootmark{a}$     & $0\tablefootmark{a}$   \\
                  &            & Luminosity ($10^{34}$ \text{W})           &	$	3.9	\pm	0.2	$	&	$	4.3	\pm	0.2	$	&	$	4.3	\pm	0.3	$ \\
%
\hline
Cr K$\alpha$ & {\tt file\tablefootmark{b}}   & nr of Ph ($10^{47} \rm ph s^{-1}$)  &	$	3.3	\pm	1.2	$	&	$	3.1	\pm	1.2	$	&	$	3.1	\pm	1.2	$ \\ 
              & {\vgau} & $\sigma$ ($\rm km/s$)   &	$	150	\pm	20\tablefootmark{d}	$	&	$	140	\pm	20\tablefootmark{d}	$	&	$	140	\pm	30\tablefootmark{d}	$     \\
              &              & Luminosity ($10^{32}$ \text{W})    &	$	2.8	\pm	1.0	$	&	$	2.7	\pm	1.0	$	&	$	2.7	\pm	0.5	$ \\
\hline
Ni K$\alpha$ & {\tt file\tablefootmark{b}}  & nr of Ph ($10^{47} \rm ph s^{-1}$)  & $3.4	\pm	1.1$	&$4.0	\pm	1.1$	&$4.0	\pm	1.1$ \\
              & {\vgau} & $\sigma \, (\rm km/s)$   &	$	150	\pm	20\tablefootmark{d}	$	&	$	140	\pm	20\tablefootmark{d}	$	&	$	140	\pm	30^{\tablefootmark{d}}	$     \\
              &             & Luminosity ($10^{32}$ \text{W})           & $4.1	\pm	1.3$	&$4.8	\pm	1.3$	&$4.8	\pm	1.3$  \\
\hline
Compton shoulder   & {\tt file\tablefootmark{b}}     & nr of Ph ($10^{47} \rm ph s^{-1}$)  &	$	6.9	\pm	3.3	$	&	$	6.4	\pm	3.3	$	&	$	6.4	\pm	2.7	$ \\
 & {\vgau} & $\sigma$ ($\rm km/s$)   &	$	150	\pm	20\tablefootmark{d}	$	&	$	140	\pm	20\tablefootmark{d}	$	&	$	140	\pm	30\tablefootmark{d}	$     \\
                            & {\tt reds}    &$\mathrm{c}*z \,(\mathrm{km/s})$      &	$6	\pm	10\tablefootmark{d}	$	&$5 \pm 10\tablefootmark{d}	$	&$	5	\pm	10\tablefootmark{d}	$       \\
     & {\tt vcom}        & skewness     & $-1.0^{+0.6}_{-0.0}$  & $-1.0 ^{+0.4}_{-0.0} $   & $-1.0 ^{+0.4}_{-0.0} $ \\     
                     
                     &             & Luminosity ($10^{32}$ \text{W})    &	$	7.2	\pm	3.4	$	&	$	6.7	\pm	3.4	$	&	$	6.7	\pm	1.3	$ \\
\hline 
                        \multicolumn{3}{c}{C-stat / expected C-stat = }   & $2046/2013$   & $2044/2013$     & $2039/2013$  \\
\hline
\hline 
\end{tabular}
\end{center}
\tablefoot{
The parameter uncertainties are given at $68\%$ confidence. 
The luminosity of each model component is calculated between 2$-$10 keV. 
\tablefootmark{a} The parameter value was fixed. 
\tablefootmark{b} All neutral line profiles taken from \cite{Holzer1997_Kab}. The norm in $\rm ph s^{-1}$ of the {\tt file} model represents the total number of photons. 
\tablefootmark{c} The inner radius of the disk is degenerate with the spin of the black hole, see text.
\tablefootmark{d} The values are tied to these for $\rm Fe \, K\alpha \beta_{nar}$.
}
\end{table*}

%
\begin{table}[!tbp]
\begin{minipage}[t]{\hsize}
\setlength{\extrarowheight}{1pt}
\setlength{\tabcolsep}{4pt}
\caption{ 
To investigate the parameters degeneracy of the best-fit {\tt spei} models shown in Table \ref{tab_best-fit-par}, we perform the grid search for the parameters of inner radius ($\rm r_{in}$), emissivity slope ($q$), spin ($a$) and inclination angle ($i$).
}
\label{table_grid_search}
\centering
\small
\renewcommand{\footnoterule}{}
\begin{tabular}{c | c c c c c c}
\hline \hline
       & $a$  & $\rm r_{in}$ / $\rm r_{isco}$     & $i$  & $q$  \\
\hline
range  & 0 $-$ 0.998  & 1 – 5   & 15.0 – 48.0  & 2.0 – 4.5 \\
\hline
step  &   0.05 &  0.05  &  1.65 &  0.125 \\

\hline
\end{tabular}
\end{minipage}
\tablefoot{ 
Given the spin, the $\rm r_{in}$ is associated with the $\rm r_{in} / r_{isco}$, which varies starting from the corresponding ISCO radius extend to the factor of 5 with step of 0.05.  
The parameter $i$ and $q$ have a step number of 20 ranging from their start and end values, respectively.
}
\vspace{0.0cm}
\end{table}

\section{Results}\label{result}

\begin{figure}[!tbp]
\centering
\hspace*{-0.00cm}\resizebox{1.0\hsize}{!}{\includegraphics[angle=0]{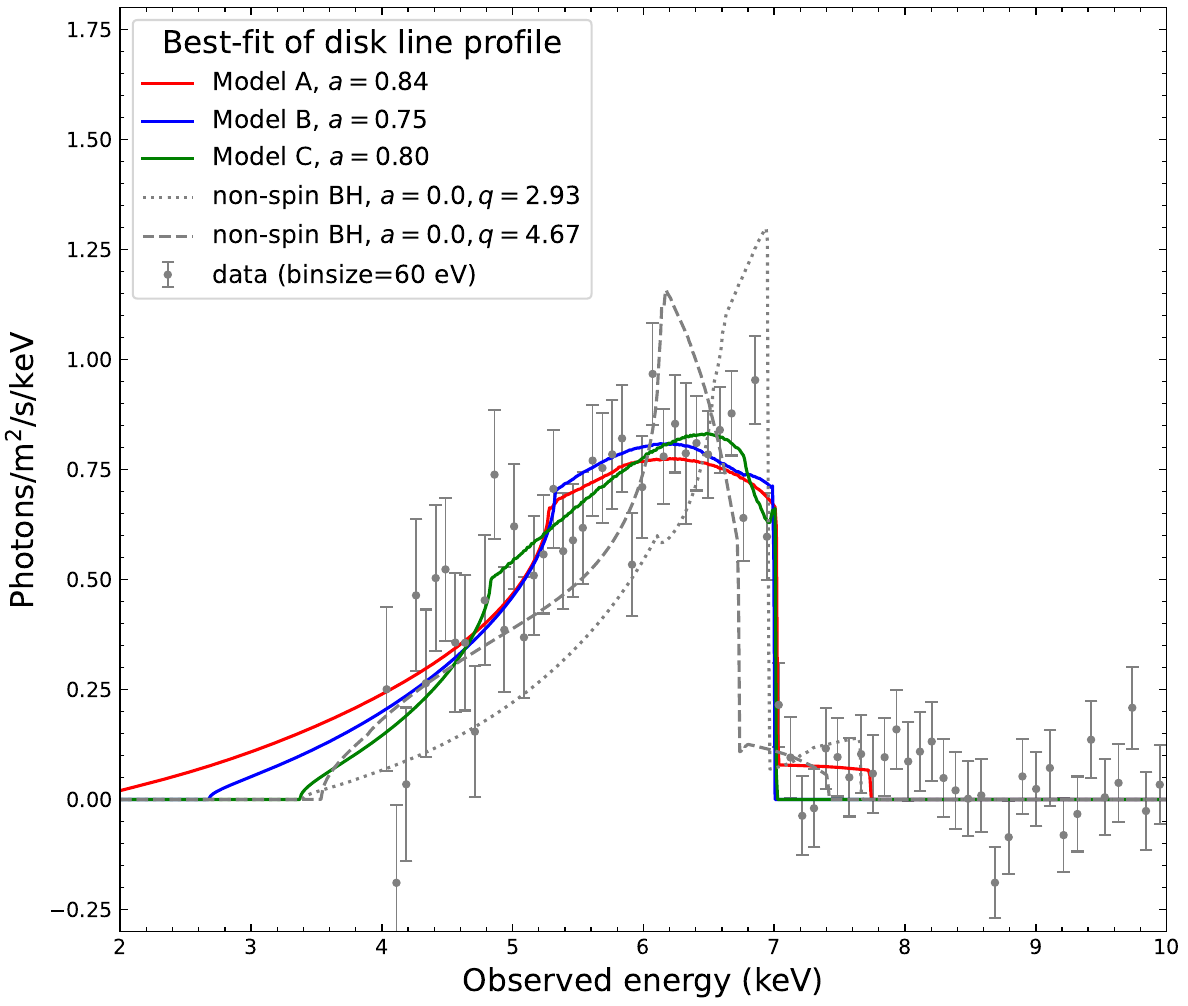}}\vspace{-0.0cm}
\caption{
Modeling relativistic iron line profiles.
The red, blue and green curves represent the best-fit relativistic iron line profiles of Model A, B and C, respectively.
The gray dotted line and dashed line are non-spin BH solutions with different emissivity slope.
The former one is derived from a grid search based on Table \ref{table_grid_search}, the later one is one of solution of Figure \ref{fig_spin_c-stat}.
Both profiles correspond to the minimum C-statistic value for a non-spinning BH case.
The gray data points represent the residuals of Model A (Data $-$ Model A) excluding the relativistic iron line, binned to 60 eV.
}
\label{fig_spei_three_models}
\end{figure}

The Resolve spectrum reveals narrow \FeKa and K$\beta$ lines, alongside a series of absorption lines from highly ionized warm absorbers. 
A broad emission structure with a $10\%$ excess above the continuum was clearly detected and is presented in the bottom panel of Figure \ref{fig_chi}. Figure \ref{fig_spec} illustrates one of the best-fit models, Model A, along with the profiles of its individual components.
Model A provides acceptable fits to the Resolve data, with a C-statistic of approximately 2046, compared to an expected value of 2013 $\pm $ 64. Additionally, the broad emission feature can be well reproduced by a relativistic Fe line component. 
We examine three different models to fit this feature, as described in Section \ref{model_bFe}. 
All these solutions yield profiles that are similar to each other, and the resulting best-fits of the three models (Model A, B, and C) are generally in agreement.
Figure \ref{fig_chi_comparison} compares the three models with the data. 
While they generally agree, a modest discrepancy appears in the narrow \FeKb feature near $\sim$ 7 keV, which is attributable to relativistic \FeKb emission added in model A. Figure \ref{fig_spei_three_models} presents the relativistic iron line profiles corresponding to the best-fits of the different models, showing that the profiles (red, blue, and green curves) differ only slightly.
Table \ref{tab_best-fit-par} presents the best-fitting parameters with 1$\sigma$ errors and C-statistic values for Models A, B, and C.

When we perform error analysis, we find several strong correlations between the parameters of the Speith profile. To investigate this, we perform a grid search to examine the degeneracy of {\tt spei} model parameters using the step command in SPEX. The step sizes and parameter ranges are shown in Table \ref{table_grid_search}. The normalization of the line is always adjusted to the best-fit value for the given parameter set.
The results of this grid search are shown in Figure \ref{fig_spin_factor_isco}, which presents the degeneracy between $\rm r_{in}/r_{isco}$ and spin. Following Figure \ref{fig_spin_factor_isco}, Figure \ref{fig_spin_rin} shows the inner radius versus black hole spin, illustrating the uncertainty in the inner radius for our current modeling.

To better constrain the spin, we fix the inner radius to the ISCO at each spin value ($r_{\rm in}=r_{\rm ISCO}$) and allow the emissivity index and inclination to vary freely (no grid limits). The resulting distribution of C-statistic values is shown in Figure \ref{fig_spin_c-stat}; the best-fitting spin and its $3\sigma$ error bar are indicated, disfavouring a non-rotating black hole.

\section{Discussion}
\label{sect_discuss}

The $439 \, \rm ks$ time-averaged spectrum of NGC 3783, taken by XRISM/Resolve in July 2024, has shown rich results.
The \FeKa emission complex is clearly resolved into three components: a narrow feature with $\rm FWHM \sim 300 \, \rm km/s$, an intermediate-width component with $\rm FWHM \sim 3500 \, \rm km/s$, and a broadest emission structure spanning $4 - 7$ keV.
Their respective equivalent widths are $EW_{\rm nar} = 48 \pm 3 \rm \, eV$,  $EW_{\rm intm} = 48 \pm 4 \rm \, eV$, and $EW_{\rm brd} = 278 \pm 14 \rm \, eV$.
The narrow \FeKa is accompanied by a weak and moderately broadened Compton shoulder. Additional weak emission lines, including Cr K$\alpha$ and Ni~K$\alpha$, were detected at a significance level of approximately $\sim 3 - 4 \sigma$, with Gaussian velocity widths comparable to the $\rm Fe \, K\alpha_{nar}$ component.

\subsection{Narrow Iron K$\alpha$ Lines}

The detection of the narrow \FeKa line component helps us to better interpret the distant reflection regions in NGC 3783.
The fluorescence line properties can potentially be used to constrain the geometry, column density, and kinematics of the material in which the line is formed (\citealp{Murphy_Yaqoob_2009}; \citealp{Yaqoob2011_Feka}; \citealp{Yaqoob2024}; \citealp{Vander_Meulen_2024}).
The spectrum shows a clear split of Fe K$\alpha1$ and Fe K$\alpha2$, which is consistent with the neutral \FeKa profile.

The FWHM of the \FeKa core of $1720 \pm 360 \, \rm km s^{-1}$ measured by the Chandra grating data from 2001 (\citealp{Kaspi2002}), lies between the narrow (FWHM = $350 \pm 50$ km/s) and intermediate (FWHM = $3510 \pm 470$ km/s) components in our analysis. 
Fitting the Resolve spectrum with a single \FeKa core component by {\tt delt*vgau} yields a FWHM of $\sim$ $1390 \pm 40$ km/s, consistent with the Chandra value.
The Chandra/HETG spectral resolution is worse than that of XRISM/Resolve at these energies by at least a factor of $\sim$ 6.
This suggests that the profile of the Fe core has remained stable over a timescale of approximately 24 years.

If we assume virial motion for the emitting clouds producing the narrow and intermediate velocity component, we have
\begin{equation}\label{eq1}
    r_{\rm Fe} = \frac{GM_{\rm BH}}{v^2}  
\end{equation}
where $M_{\rm BH}$ is the black hole mass, $v$ is the de-projected orbital velocity. Following \citet{Gandhi2015}, we relate $v$ to the observed line width and the line-of-sight inclination 
$i$ as  
\begin{equation}\label{eq2}
    v = \frac{v_{\rm FWHM}}{sin(i)}
\end{equation}
where $v_{\rm FWHM} = 2.35 \sigma_v$, the $\sigma_v$ is the Gaussian width of the line.
Equations \ref{eq1}–\ref{eq2} imply 
\begin{equation}\label{eq3}
    r_{\rm Fe} = \frac{GM_{\rm BH} sin^2(i)}{5.5 \sigma_v^2}
\end{equation}.
Applying Eq. \ref{eq3}, we adopt $i \approx 23$ degree \citep{GRAVITY_Collaboration_2021_central_parsec}, and $\rm M_{BH} = 2.54^{+0.90}_{-0.72} \times 10^{7} M_{sun}$ for NGC 3783, which is estimated based upon a joint analysis that combines both GRAVITY and optical reverberation mapping data \citep{GRAVITY_Collaboration_2021_geometric_distance} to estimate the distance of \FeKa complex.

The narrow component is likely located at $\sim 5 \times 10^5$ $\rm R_g$, corresponding to $\sim 0.64$ pc with a typical $\sim 30\%$ uncertainty. This is a factor $\sim$ 4–5, beyond the hot-dust radius ($\sim 0.14$ pc) measured with near-IR interferometry \citep{GRAVITY_Collaboration_2021_central_parsec}, in contrast to \citet{Gandhi2015}, who inferred an origin inside the hot-dust region from Chandra/HETGS data.
Given $r_{\rm Fe} \propto sin^2(i)$ dependence, the estimated $r_{\rm Fe}$ is sensitive to the line-of-sight inclination. Accounting for the uncertainty in $i$, the inferred $r_{\rm Fe}$ remains broadly consistent with the results of \cite{Gandhi2015}.
Moreover, geometry can further bias the effective viewing angle: if our sight-line is dominated by the illuminated inner face of the torus/BLR, the effective inclination is smaller with big opening angle.

The intermediate Fe core component is likely emitted from a radius of $\sim 5 \times 10^3$ Rg ($\sim$ 0.006 pc or 7.6 light-days), which falls within the broad-line region ($\sim$ 3–20 light-days) reported from optical spectroscopy by \cite{Bentz202101a} and \cite{Bentz202110b}.
In their work, the inferred kinematics of the outer BLR as probed by $\rm H \beta$ are dominated by near-circular orbits with a contribution from infall, whereas the kinematics of the inner BLR probed by \ion{He}{II} is dominated by an unbound outflow.
The difference in kinematics between the $\rm H \beta$ and \ion{He}{II} emitting regions of the BLR is intriguing given the large changes in the ionizing luminosity of NGC 3783 during an obscured state \citep{Mehdipour2017} and evidence for possible changes in the BLR structure as a result.
A caveat in comparing the XRISM/Resolve measurements with BLR sizes from earlier optical campaigns is the source state: those optical data were taken when NGC 3783 was obscured, whereas our 2024 XRISM observations captured it in an unobscured state. A coordinated XRISM campaign with simultaneous optical spectroscopy would be the most robust way to test whether the intermediate \FeKa core is physically connected to the BLR.

In Table \ref{tab_best-fit-par}, the centroid of the Fe K$\alpha$ complex 
for both narrow and intermediate-width were found to be within 10 and 300 km/s redshift 
from their rest frame energy, respectively.
These line centroids are within the systematic uncertainty of the Resolve instrument.
Previously, an excess redshift velocity of $620^{+80}_{-70}$ km/s of the iron K$\alpha$ line with respect to the host galaxy was found in NGC 3783 using Chandra ACIS-S/HETGS (\citealp{Danehkar2025}), which implies more redward photons were observed in the far side of the disk, potentially caused by a warped disk \citep{Miller2018}.
The BLR in NGC 3783 can be accurately characterized as a thick, rotating disk with the highest cloud concentration in the inner region by \cite{GRAVITY_Collaboration_2021_central_parsec}.
The excess redshift of the \FeKa\ emission found by \cite{Danehkar2025} could stem from a transient event, such as X-ray emitting BLR inflow towards the supermassive black hole, or a failed wind falling back towards the inner region (\citealp{Mehdipour2017}; \citealp{Li2025}).

Recent XRISM spectroscopic observations of NGC 4151 also suggest contributions from the innermost BLR (optical/X-ray) to the \FeKa line, in addition to a potentially warped accretion disk \citep{XRISM_collaboration2024}.
Alternatively, as proposed by \citealp{Miller2018}, a failed wind could also result in an asymmetric narrow \FeKa line.

\subsection{Chromium and Nickel lines}
In addition to Fe, Cr K$\alpha$ and Ni K$\alpha$ lines are detected with significances of $2.8\sigma$ and $3 - 4 \sigma$, respectively. 
The observed line broadening is consistent with the velocity dispersion of the narrow Fe core.
If we assume that the line flux for element Z scales as $\sim I_{z} \omega_{z}A_{z}$, where $I_z$ is the K-shell photoionization rate (ionisation/ion), $\omega_z$ is the fluorescence yield and $A_z$ is the abundance relative to hydrogen, we can estimate the expected line fluxes for Cr and Ni by scaling these parameters to the observed Fe flux.
This yields predicted fluxes of $1.3 \times 10^{47} \rm photons \, s^{-1}$ for Cr and $4.0 \times 10^{47} \rm photons \, s^{-1}$ for Ni, which are in broad agreement with the observed values of $3.3 \pm 1.2$ and $3.4 \pm 1.1$ $\times$ $10^{47} \rm photons \, s^{-1}$.
The lines are therefore consistent with solar abundance ratios, although Cr might also be slightly enhanced. 
We need better data, e.g. a larger effective area (gate valve open) or longer exposure time with Resolve, to perform accurate abundance measurement.
%

\subsection{The Compton shoulder of \rm Fe K$\alpha_{\rm nar}$}

The potential Compton shoulder of the Fe K$\alpha$ core is marginally detected at approximately $2.4 \sigma$ confidence. Its shape remains largely unconstrained (see Figure \ref{fig_cs_chi}), and its flux is estimated to be no more than $10\%$ of the narrow \FeKa line.
This indicates an optical depth $\tau = 0.1$, corresponding to a column density of matter emitting the Compton shoulder of $\sim 10^{23} \rm m^{-2}$ associated with scattering material.
The luminosity of the narrow component of \FeKa is close to that of the intermediate one (see Table \ref{tab_best-fit-par}). 
Modelling shows that we cannot distinguish between the narrow and intermediate component as the source of the scattered photons.

In contrast to the results reported by \cite{Yaqoob2005}, where the broad Fe line and Compton shoulder scenarios were degenerate in the Chandra data, our XRISM observations now allow us to clearly distinguish between them.

\subsection{Relativistic reflection emission}\label{result_broad_Feka}

%
\begin{figure}[!tbp]
\centering
\hspace*{-0.0cm}\resizebox{1.10\hsize}{!}{\includegraphics[angle=0]{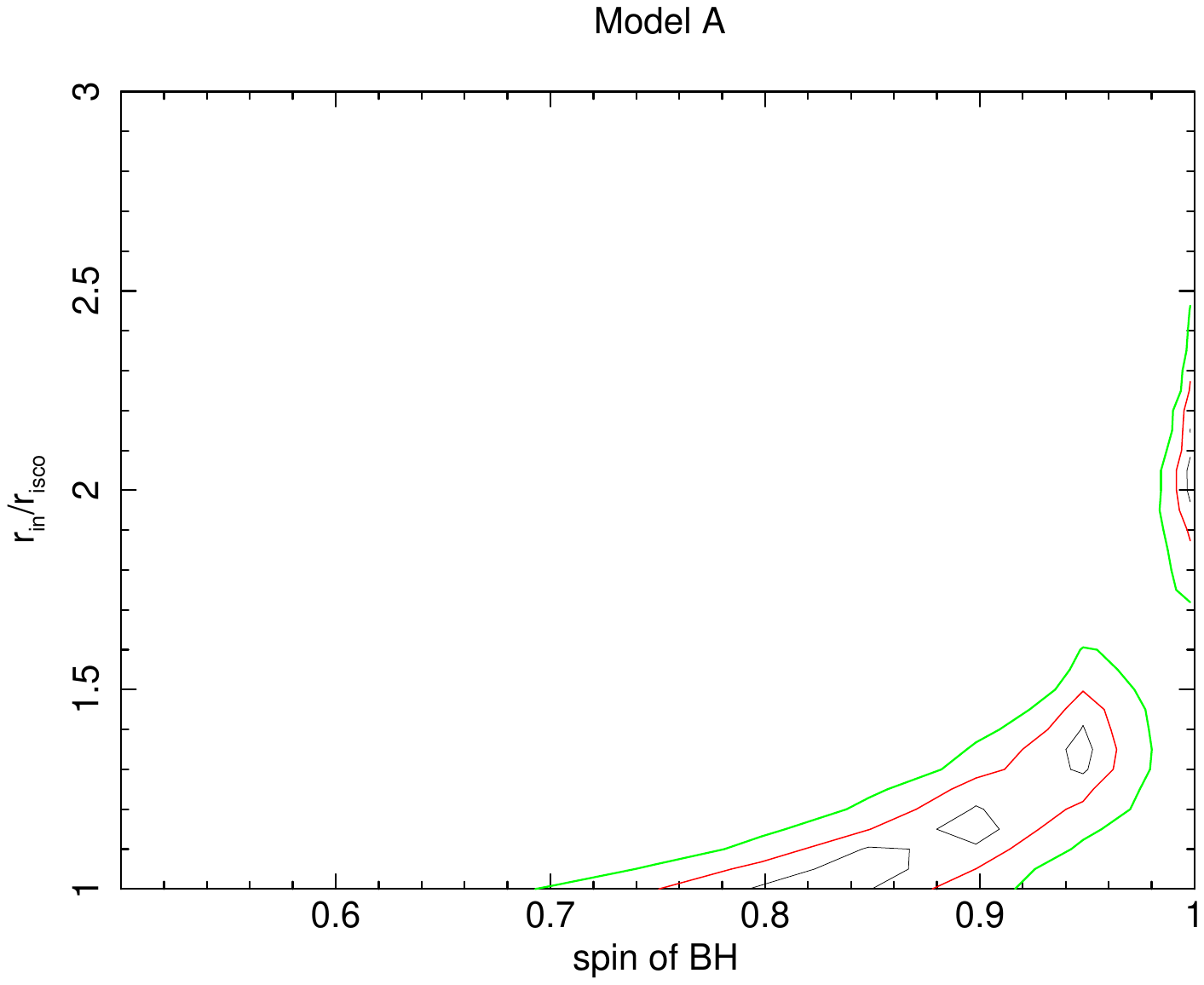}}\vspace{-0.1cm}
\hspace*{-0.0cm}\resizebox{1.10\hsize}{!}{\includegraphics[angle=0]{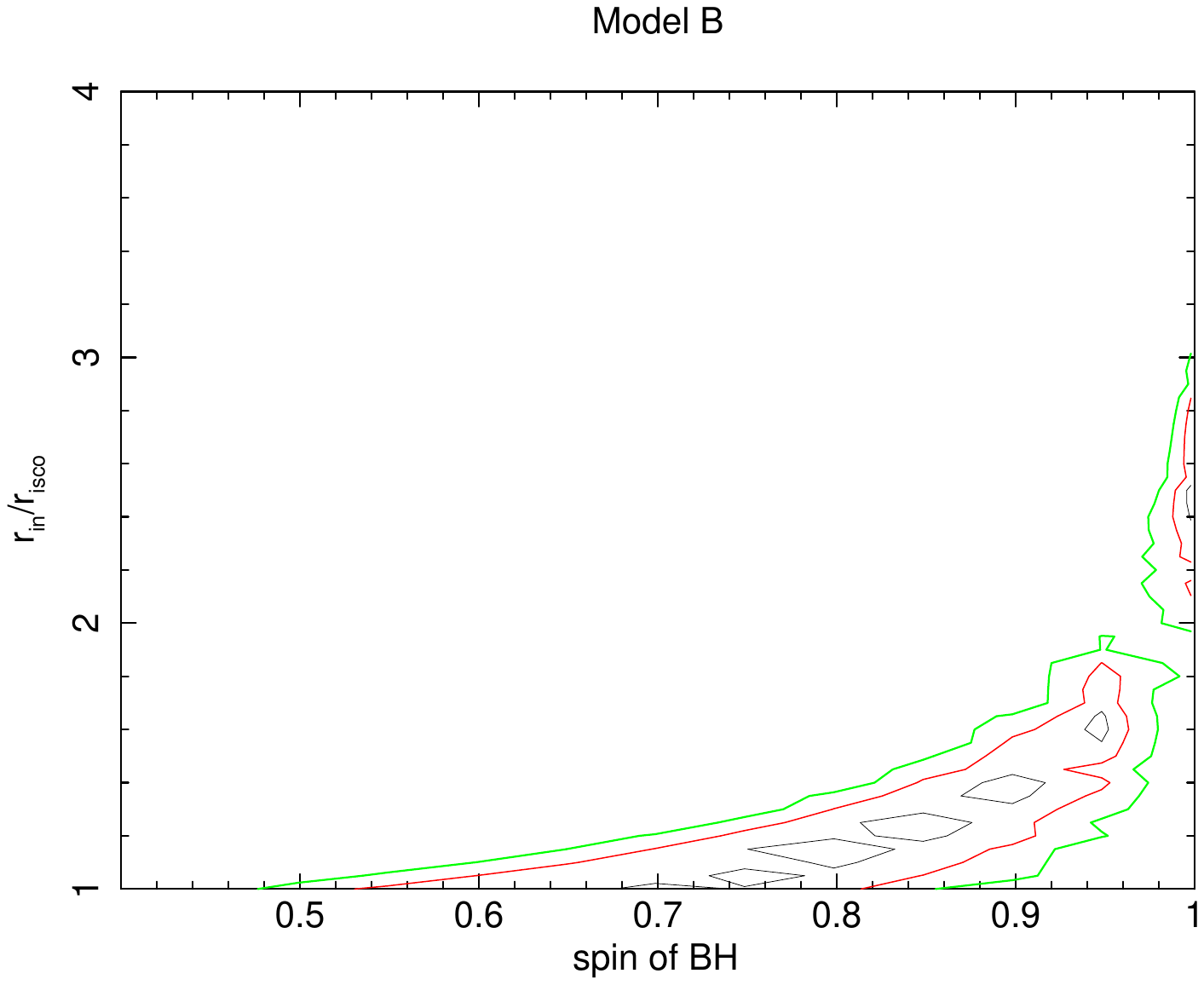}}\vspace{-0.1cm}
\hspace*{-0.0cm}\resizebox{1.10\hsize}{!}{\includegraphics[angle=0]{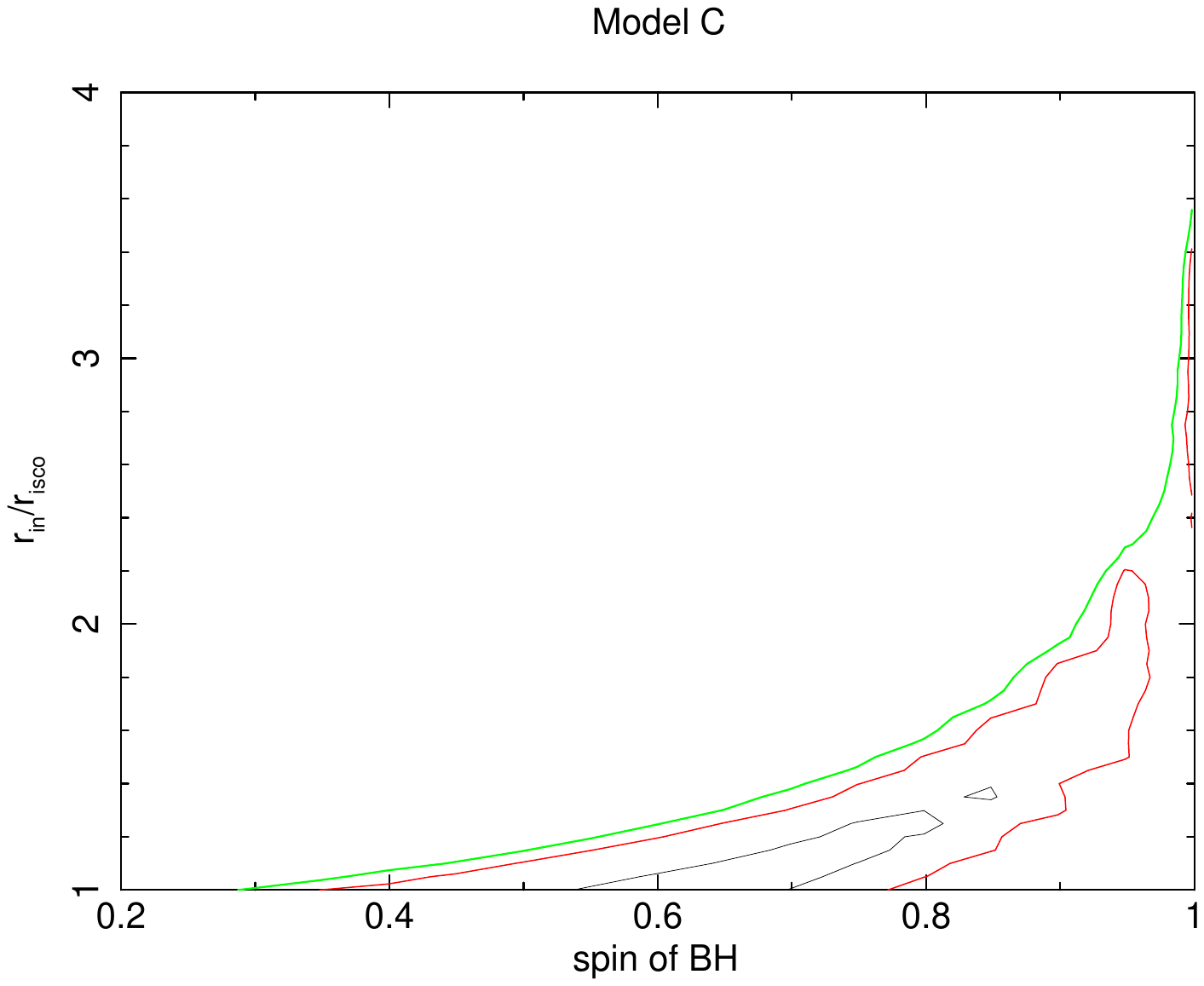}}
\caption{
The contour plot between the ratio of inner radius to ISCO ($\rm r_{in} / r_{isco}$) and spin of BH with Resolve time-averaged spectrum in NGC 3783.
The same parameter space is applied for the grid search of Model A, B and C.
The $1\sigma$ (68$\%$), $2\sigma$ (95$\%$) and $3\sigma$ (99.7$\%$) confidence contour are in black, red and light green, respectively.
}
\label{fig_spin_factor_isco}
\end{figure}

\begin{figure}[!tbp]
\centering
\hspace*{-0.0cm}\resizebox{1.0\hsize}{!}{\includegraphics[angle=0]{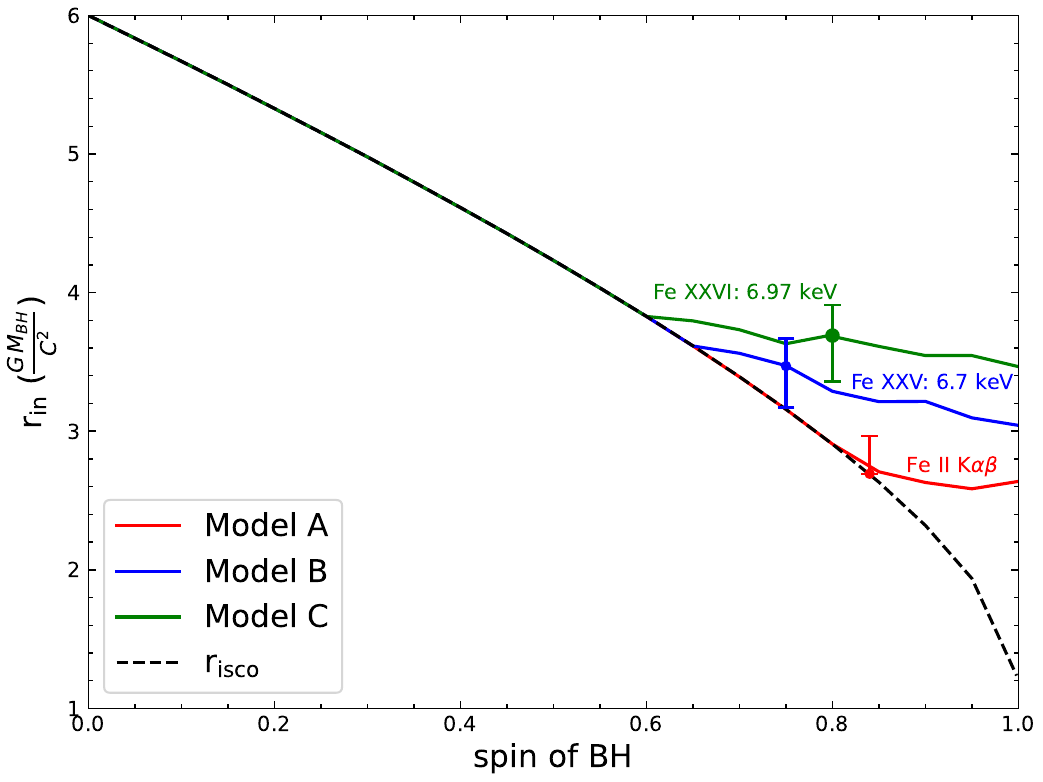}}\vspace{0.1cm}
\caption{
The best-fit inner radius as function of spin is shown by the coloured lines.  
For each model, the best-fit inner radius is shown by a filled circle, together with the associated $1\sigma$ uncertainty from Table \ref{tab_best-fit-par}.
The prograde ISCO radius as a function of spin is shown as the black dashed line.
}
\label{fig_spin_rin}
\end{figure}

\begin{figure}[!tbp]
\centering
\hspace*{-0.0cm}\resizebox{\hsize}{!}{\includegraphics[angle=0]{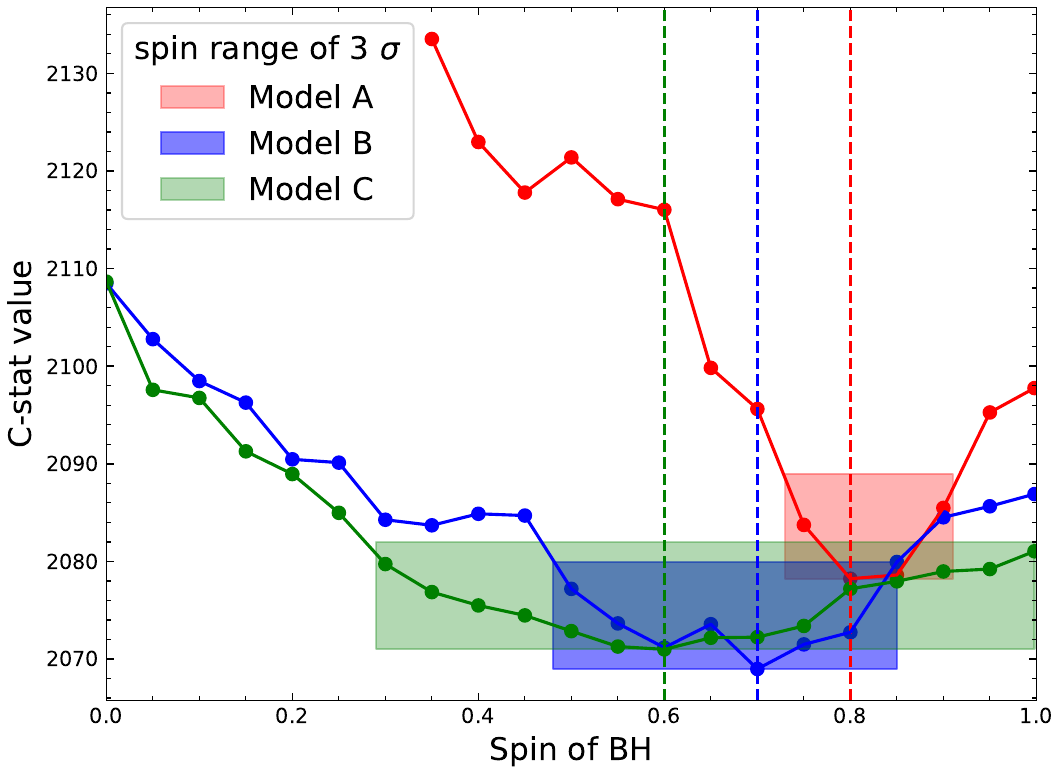}}\vspace{0.1cm}
\caption{
Distribution of C-statistic values as a function of spin, with the inner radius fixed to the ISCO at each spin ($r_{\rm in}=r_{\rm ISCO}$).
For each model, the best-fitting spin is indicated by a vertical line.
We show the $3\sigma$ confidence interval on the spin (shaded coloured region), clearly demonstrating that a non-spinning black-hole solution is ruled out.
}
\label{fig_spin_c-stat}
\end{figure}

One of the derived parameter from our model is the inclination angle of the accretion disk. A known degeneracy exists between the inclination, which is primarily determined by the blue edge of the relativistic line profile, and the intrinsic energy of the line. 
This is seen in Table \ref{tab_best-fit-par}, where the inclination angle is found to decrease from $\sim 44^\circ$, to $\sim 32^\circ$, $\sim 17^\circ$  with increasing ionization state (neutral iron, \ion{Fe}{XXV} and \ion{Fe}{XXVI}, respectively).
The emissivity slope of the disk is found to be in the range of 3 – 4. No clear correlation is found between this slope and other parameters, such as the ionization state and the inclination angle.

The inner disk radius is a free parameter in our analysis. As shown in Figure \ref{fig_spin_factor_isco}, this approach reveals a degeneracy between the ratio of the inner radius to ISCO and the black hole spin i.e. as the spin increases, the inner radius moves away from the ISCO.
The black $1 \sigma$ contour in Fig. \ref{fig_spin_factor_isco} indicates that different spin solutions are possible in the disk line model.
Based on Table \ref{table_grid_search}, our grid search also rules out non-spinning solutions with at least 3 $\sigma$ significance.
The observed spectrum can be fitted equally well by two scenarios: a high-spin black hole with a potentially truncated disk, or a mildly-spinning black hole with a disk extending to the ISCO inner radius.

Following up of Fig. \ref{fig_spin_factor_isco}, we examine the potential relationship between the absolute value of the inner radius and the black hole spin. As shown in Figure \ref{fig_spin_rin}, while the ISCO radius decreases sharply with increasing spin (black dash line), the observed inner radius exhibits only minor variation for different spin values. The best-fit inner radius for the neutral line model is approximately 2.5 $-$ 3 $\rm R_{g}$. This value increases to 3 $-$ 3.5 $\rm R_{g}$ for the ionized line in Model B and then slightly increases to 3.5 $-$ 4 $\rm R_{g}$ for Model C. 
Since the observed profile likely blends emission from neutral, He-like, and H-like iron, we fit each ionization state separately and find a consistent result: the best-fit inner radius is $\rm r_{in} \approx 2.5$ $-$ $4\rm  R_{g}$ in all case, where best-fit spin $a > 0.6$ corresponding to $\rm r_{ISCO} < 4 R_{g}$, which is strongly dependent on spin.

We now focus on the scenario that the disk is not truncated but extends to the ISCO radius. This possibility is not excluded from our analysis.
To get proper constraints on the spin, we refit data with fixing $\rm r_{in} = r_{ISCO}$ for each spin, as described in Sect. \ref{result}. 
The C-statistic value versus spin is plotted in Figure \ref{fig_spin_c-stat}.
The fitting results indicate that the allowed spin range for Model A, B and C are $0.73 \le a \le 0.91$, $0.48 \le a \le 0.85$ and $0.29 \le a \le 0.998$ with $3 \sigma$ confidence level, respectively.
Additionally, a non-spinning BH case leads to a significantly worse fit, with the C-stat increasing by 168, 70 and 42 for Model A, B and C, respectively.
These results indicate that a non-spinning solution is ruled out.
The non-spinning BH model profile for Model A is shown as a dotted line in Figure \ref{fig_spei_three_models}.

We further examined the case of a non-spinning black hole with a high emissivity slope. We fixed a = 0 for Model A, while varying q from 2.0 to 6.0. The new best-fit yields a C-statistic value of 2187, which is worse than the best fit value of Model A. This indicates an emissivity slope of $q = 4.67^{+0.28}_{-0.63}$ and an inclination of $i = 34.11^{+0.26}_{-1.23}$ degrees for the disk profile (see gray dashed line in Figure \ref{fig_spei_three_models}). 
This result also rules out the non-spinning black hole scenario with a high emissivity slope based on the C-statistic value.

\subsection{Comparison with previous studies and systematic uncertainty}

Our derived inclination angle of $17.6 \pm 1.4$ degrees from Model C is slightly lower than the previous values: 19 – 22 degrees \citep{Brenneman2011} and 22 – 24 degree \citep{Reynolds2012}.
A rotating BLR model fitted to the VLTI/
GRAVITY observations of NGC 3783 implies an inclination angle of the disk of
either $23^{+16}_{-10}$ degrees \citep{GRAVITY_Collaboration_2021_central_parsec} or 
$32 \pm 4 $ degrees when combined with a reverberation mapping analysis  \citep{GRAVITY_Collaboration_2021_geometric_distance}. This is in agreement with the confidence limits of our values obtained for the inclination angle of the accretion disk by Model B ($32 \pm 2$ degree).

As shown in Table \ref{tab_best-fit-par}, the emissivity slope $q$ falls within the range of 3 $-$ 4 across the three models. This result aligns with the standard emissivity profile expected in flat spacetime. 
It differs from the steeper profiles ($q \sim 5 - 6$) observed in some AGNs (such as MCG$-$6-30-15, \cite{Fabian2003}) due to strong GR light blending depending on the location/geometry of an illuminating source above the disk (\citealp{Fukumura2007}; \citealp{Dauser2013}; \citealp{Wilkins_Fabian_2012}).

All of our best-fit models prefer spinning BH solutions as seen in Figure \ref{fig_spin_factor_isco} and Figure \ref{fig_spin_rin}.
Our derived lower limit for the spin is $a \ge 0.29$ at the $3\sigma$ significance level as seen in Figure \ref{fig_spin_c-stat}.
It is consistent with values reported by \cite{Brenneman2011} ($a \ge 0.88$ with $99\%$ confidence), \cite{Reynolds2012} (a rapidly spinning black hole $a > 0.89$) and \cite{Danehkar2025} (a near-maximal SMBH spin $a = 0.98^{+0.02}_{-0.12}$).
The large uncertainty on spin measurement in our analysis compared to other results can be due to many causes.

The measured spin here should be viewed as an initial constraint, with a more thorough analysis to be performed that includes the entire reflection spectrum rather than just the \FeKa line.  The latter approach will help to eliminate many of the model degeneracies found in this paper (e.g., the ionization state of the disk and the energy of the line, which impacts the measurement of the blue wing and inclination).

Excluding the absorption-line data points between 6.38 $-$ 7.1 keV has no impact on our current final results.
In principle, many factors affect the modeling of the relativistic iron line component. This includes the accounting for the intermediate-width Fe K$\alpha$$\beta$ component, inclusion of a relativistic \FeKb component, the absorption trough modelings of the \ion{Fe}{XXVI} line, and the modeling of the remaining \ion{Fe}{XXVI} emission line.
This is hard or impossible to do with CCD spectra or grating spectra with relatively low sensitivity.
While the Resolve instrument offers the spectral resolving power needed to precisely characterize the narrow emission and absorption features in the Fe-K region, it does not provide the same level of data statistics as the current CCD instruments utilized in several previous studies for the broad relativistic iron line component. A joint analysis incorporating the Xtend data would be ideal to further improve the constraints on the black hole spin parameter.

A second difference between previous work and ours is 
that we use a single emission line while others use a full reflection model such as {\tt relxill} \citep{Dauser2016} to characterise the emission. 
Although such reflection models are dominated by one strong spectral line for a given ionization state of the accretion disk in the 4 $-$ 10 keV band of the Resolve spectrum, a self-consistency reflection model combining with soft and hard X-ray data will further improve spin measurement.

Finally, the spin measurement may be biased if the relativistic line shows strong variability. Time variability is the subject of follow-up papers, but a simplified preliminary analysis shows that apart from a strong soft flare (see Gu et al. accepted, paper III), the equivalent width of the relativistic line varies by no more than 30\%. Therefore, variability is expected to play a minor role in our spin determination.

\subsection{Open question and future work}
The XRISM data alone cannot strongly constrain the black hole spin, especially when the inner radius of the disk is left as a free parameter. The Compton hump observed in NuSTAR data has the same origin as the Fe line and could help alleviate some of the model degeneracies. The lower energy bound of 4 keV likely also limits our ability to make robust spin constraints. Many previous spin measurements using reflection spectra were driven by the shape of the soft excess, which contains significantly more photons than the iron line. However, whether the soft excess is dominated by disk reflection or warm corona emission remains an active area of debate.
Our approach serves as a first step before more physically motivated, self-consistent models are employed. Combining soft X-ray data from XMM-Newton characterizing the soft X-ray excess with hard X-ray observations from NuSTAR to constrain the Compton hump, together with our Resolve results, will provide a more comprehensive understanding of the accretion flow and \FeKa emission line physics.

\begin{acknowledgements}
We thank the anonymous referee for his/her constructive comments.
C.L. acknowledges support from the Chinese Scholarship Council (CSC) and Leiden University/Leiden Observatory, as well as SRON.
SRON is supported financially by NWO, the Netherlands Organization for Scientific Research.
%
Part of this work was performed under the auspices of the U.S. Department of Energy by Lawrence Livermore National Laboratory under Contract DE-AC52-07NA27344.
M. Mehdipour acknowledges support from NASA grants 80NSSC23K0995 and 80NSSC25K7126.
%
M.Signorini acknowledges support through the European Space Agency (ESA) Research Fellowship Programme in Space Science.
C.L. thanks Bert Vander Meulen, Elisa Costantini, Anna Jur\'{a}\u{n}ov\'{a} for the helpful discussions. 
\end{acknowledgements}

\bibliographystyle{aa}
\bibliography{references}

\end{document}